\documentstyle[12pt]{article}
\begin{document}

\tolerance=5000

\def\pp{{\, \mid \hskip -1.5mm =}}
\def\cL{{\cal L}}
\def\be{\begin{equation}}
\def\ee{\end{equation}}
\def\bea{\begin{eqnarray}}
\def\eea{\end{eqnarray}}
\def\tr{{\rm tr}\, }
\def\nn{\nonumber \\}
\def\e{{\rm e}}
\def\D{{D \hskip -3mm /\,}}

\  \hfill 
\begin{minipage}{3.5cm}
OCHA-PP-139 \\
NDA-FP-?? \\
August 1999 \\
\end{minipage}

\vfill

\begin{center}
{\large\bf AXION-DILATONIC CONFORMAL ANOMALY FROM 
AdS/CFT CORRESPONDENCE}

\vfill

{\sc Shin'ichi NOJIRI}\footnote{nojiri@cc.nda.ac.jp}, 
{\sc Sergei D. ODINTSOV}$^{\spadesuit}$\footnote{
odintsov@mail.tomsknet.ru}, \\
{\sc Sachiko OGUSHI}$^{\heartsuit}$\footnote{
JSPS Research Fellow, 
g9970503@edu.cc.ocha.ac.jp}, 
{\sc Akio SUGAMOTO}$^{\heartsuit}$\footnote{
sugamoto@phys.ocha.ac.jp}, \\
and {\sc Miho YAMAMOTO}$^{\heartsuit}$\footnote{
g9840530@edu.cc.ocha.ac.jp}

\vfill

{\sl Department of Mathematics and Physics \\
National Defence Academy, 
Hashirimizu Yokosuka 239, JAPAN}

\vfill

{\sl $\spadesuit$ 
Tomsk Pedagogical University, 634041 Tomsk, RUSSIA}

\vfill

{\sl $\heartsuit$ Department of Physics, 
Ochanomizu University \\
Otsuka, Bunkyou-ku Tokyo 112, JAPAN}

\vfill

{\bf ABSTRACT}

\end{center}

We discuss general multidimensional axion-dilatonic AdS gravity 
which may correspond to bosonic sector of Gibbons-Green-Perry 
(compactified) IIB supergravity with RR-scalar (axion). Using 
AdS/CFT correspondence the 4d conformal anomaly on 
axion-dilaton-gravitational background is found from SG side. 
It is shown that for IIB SG with axion such conformal anomaly 
coincides with the one obtained from QFT calculation 
in ${\cal N}=4$ super Yang-Mills theory conformally coupled with 
${\cal N}=4$ conformal SG. Brief discussion on possibility to apply 
these results for gauged SGs is also presented.

\newpage


One of the important tests of AdS/CFT correspondence (for an 
introduction and review, see \cite{AGMOO}) is SG side calculation of 
conformal anomaly also called Weyl or trace anomaly (for a review, 
see \cite{Duff}). This may also help in the construction of new 
examples of anomaly in the situation when perturbative (QFT) 
calculations are extremely complicated. The method to make such 
calculation (from SG side) has been suggested in ref.\cite{1}. 
This evaluation has been tested for ${\cal N}=4$ super YM theory 
in refs.\cite{NOa,BK,MV,HS,MN} and for 2d superconformal 
theory \cite{NOa,NT}, mainly on pure gravitational background. 
Exact agreement with QFT conformal anomaly has been established in 
the leading order of large $N$ expansion. Moreover,
it has been found agreement \cite{BNG,NO1} between AdS/CFT 
conformal anomaly and perturbative QFT anomaly in ${\cal N}=2$ 
SCFT even in the next to leading order term.

In ref.\cite{NOa} the AdS/CFT correspondence on the level of trace 
anomaly has been successfully checked on dilaton-gravitational 
background. It has been demonstrated that boundary side is 
${\cal N}=4$ super YM theory conformally coupled with 
${\cal N}=4$ conformal SG. However, the important question 
remains: May other fields spoil AdS/CFT correspondence? 
To answer this question we consider 5d dilaton-axionic AdS gravity.
Our action is actually motivated by bosonic sector of SG model due to 
Gibbons, Green and Perry\cite{GGP}which contains axion.
This IIB SG represents the special case of our action. Other SGs 
maybe considered as well.

We found SG side conformal 
anomaly which exactly agrees with perturbative conformal anomaly in 
${\cal N}=4$ super YM theory conformally coupled with 
${\cal N}=4$ conformal supergravity. It completes the proof of
AdS/CFT correspondence  between two theories \cite{NOa}. 
It is imporatant that background axion and dilaton are kept 
in the calculation of conformal anomaly.

We start with the following action:
\begin{eqnarray}
\label{oi}
S = \frac{1}{16\pi G} \left\{ \int _{M_{d+1}} d^{d+1}x \sqrt{-\hat{G}}
({\hat R} + X(\phi ,\chi )(\hat{\nabla } \phi) ^{2} + Y(\phi ,\chi )
\hat{\Delta}\phi \right. &&  \nonumber \\
+Z(\phi ,\chi )(\hat{\nabla } \chi) ^{2} + W(\phi ,\chi )
\hat{\Delta}\chi +4\lambda ^{2} ) 
\left.+\int _{M_{d}}d^{d}x\sqrt{-\hat{g}}
(2\hat{\nabla}_{\mu}n^{\mu}+\alpha) \right \} .&&
\end{eqnarray}
Here $\phi$ and $\chi$ are dilaton and axion (or RR-scalar), 
respectively, and $M_{d+1}$ is the $d+1$ dimensional manifold whose 
boundary is the $d$ dimensional manifold $M_d$ and $n_\mu$ is the 
unit normal vector to $M_d$. 
We choose the metric $\hat G_{\mu\nu}$ on $M_{d+1}$ and 
the metric $\hat g_{\mu\nu}$ on $M_d$ in the following form
\be
\label{ii}
ds^2\equiv\hat G_{\mu\nu}dx^\mu dx^\nu 
= {l^2 \over 4}\rho^{-2}d\rho d\rho + \sum_{i=1}^d
\hat g_{ij}dx^i dx^j \ ,\quad 
\hat g_{ij}=\rho^{-1}g_{ij}\ .
\ee
Here $l$ is related with $\lambda^2$ by $4\lambda^2=-d(d-1)/l^2$.
Note that we follow to method of calculation in ref.\cite{NOa} 
where dilatonic gravity has been considered.

We expand $\phi$ and $\chi$ as power series on $\rho$:
\begin{eqnarray}
\phi &=& \phi_{(0)}+\rho\phi_{(1)}+\rho^{2}\phi_{(2)}+\cdots
 + \rho^{\frac{d}{2}}\phi_{(d/2)} -\rho ^{\frac{d}{2}}\ln \rho \psi
 +{\cal O}(\rho ^{\frac{d}{2}+1}) \\
\chi &=& \chi_{(0)}+\rho\chi_{(1)}+\rho^{2}\chi_{(2)}+\cdots 
+ \rho ^{\frac{d}{2}}\chi_{(d/2)}-\rho ^{\frac{d}{2}}\ln \rho \chi 
+{\cal O}(\rho ^{\frac{d}{2}+1}) \nn 
g_{ij} &=& g_{(0)ij}+\rho g_{(1)ij}+\rho^{2}g_{(2)ij}+\cdots +
 \rho ^{\frac{d}{2}}g_{(d/2)ij}-\rho ^{\frac{d}{2}} \ln \rho h_{ij} 
+{\cal O}(\rho ^{\frac{d}{2}+1}) \ .\nonumber
\end{eqnarray}
In the following, we abbreviate the index ``(0)'' if there 
is no any confusion and 
the action (\ref{oi}) diverges in general since the action 
contains the infinite volume integration on $M_{d+1}$. 
The action is regularized by introducing the infrared cutoff 
$\epsilon$ and replacing 
\be
\label{vi}
\int d^{d+1}x\rightarrow \int d^dx\int_\epsilon d\rho \ ,\ \ 
\int_{M_d} d^d x\Bigl(\cdots\Bigr)\rightarrow 
\int d^d x\left.\Bigl(\cdots\Bigr)\right|_{\rho=\epsilon}\ .
\ee
The subtraction of the terms proportional to the inverse power of 
$\epsilon$ does not break the invariance under the scale 
transformation $\delta g_{ \mu\nu}=2\delta\sigma g_{ \mu\nu}$ and 
$\delta\epsilon=2\delta\sigma\epsilon$ . When $d$ is even, however, 
the term proportional to $\ln\epsilon$ appears. This term is not 
invariant under the scale transformation and the subtraction of 
the $\ln\epsilon$ term breaks the invariance. The variation of the 
$\ln\epsilon$ term under the scale transformation 
is finite when $\epsilon\rightarrow 0$ and should be canceled 
by the variation of the finite term (which does not 
depend on $\epsilon$) in the action since the original action 
(\ref{oi}) is invariant under the scale transformation. 
Therefore the $\ln\epsilon$ term $S_{\rm ln}$ gives the Weyl 
anomaly $T$ of the action renormalized by the subtraction of 
the terms which diverge when $\epsilon\rightarrow 0$ by 
\be
\label{vib}
S_{\rm ln}=-{1 \over 2}
\int d^4x \sqrt{-g }T\ .
\ee
Note that a bit different methods to calculate anomaly have been 
discussed in refs.\cite{MV,MN}. Nevertheless, the final result is 
unique. When $d=4$, $S_{\rm ln}$ is given by
\begin{eqnarray}
\lefteqn{S_{\ln } = \frac{1}{16\pi G} \int d^{4}x\sqrt{-g }\left\{
\frac{1}{2 \ell}g^{ij} g^{kl} 
\left(g_{(1)ij}g_{(1)kl}-g_{(1)ik}g_{(1)jl}\right)
\right. } \nn
&& +\frac{\ell}{2}\left(R ^{ij}-\frac{1}{2}g^{ij} R 
\right)g_{(1)ij} -\frac{2}{\ell}\{ (X -Y ')\phi_{(1)}\phi_{(1)} 
+(Z -W ^{*}) \chi _{(1)}\chi_{(1)} \nn
&& -(Y ^{*}+W ')\phi_{(1)} \chi_{(1)} \} 
-\frac{\ell}{2}g^{ij} \{ (X' \phi_{(1)} +X^{*} 
\chi_{(1)})\partial_{i}\phi \partial_{j}\phi  \nn
&& + \left(Z' \phi_{(1)} +Z^{*} \chi_{(1)}\right)
\partial_{i}\chi \partial_{j}\chi  \} 
-\ell g^{ij} \{X \partial_{i}\phi \partial_{j}
\phi_{(1)} +Z \partial_{i}\chi \partial_{j}\chi_{(1)} \} \nn
&& +\frac{\ell}{2}g^{ik} g^{jl} g_{(1)kl}
\{ X \partial_{i}\phi \partial_{j}\phi 
+Z \partial_{i}\chi \partial_{j}\chi \} 
-\frac{\ell}{4}g^{kl} g^{ij} g_{(1)kl}
\{ X \partial_{i}\phi \partial_{j}\phi 
+Z \partial_{i}\chi \partial_{j}\chi \}\nonumber \\
& & -\frac{\ell}{2}\left\{ (Y' \phi_{(1)}+Y^{*} 
\chi _{(1)} )\triangle \phi +(W' \phi _{(1)}+W ^{*}\chi _{(1)})
\triangle \chi + Y \triangle \phi_{(1)} +W 
\triangle \chi_{(1)} \right\} \nonumber \\
& & -\frac{\ell}{2}\left\{ Y \frac{1}{\sqrt{-g }}
\partial_{i}\left(\sqrt{-g }
(\frac{1}{2}g^{kl} g_{(1)kl}g^{ij} -g^{ik} g_{(1)kl}
g^{jl} )\partial_{j}\phi \right) \right.\nonumber \\
& & +\left.W \frac{1}{\sqrt{-g }}\partial_{i}
\left( \sqrt{-g }(\frac{1}{2}g^{kl} g_{(1)kl}g^{ij} 
-g^{ik} g_{(1)kl}g^{jl} )
\partial_{j}\chi \right) \right\} .\label{Sln} \nn
\lefteqn{\triangle \cdot \equiv 
\partial_{i}\left(\sqrt{-g }g^{ij} \partial_{j} \cdot \right)}
\end{eqnarray}
Here `` $'$ ''and `` $*$ ''express the derivative with respect 
to $\phi$ and $\chi$, respectively.
We can solve the equations of motion given by the variation of 
$S_{\ln}$ with respect to $g_{(1)ij},\phi_{(1)},\chi_{(1)}$.
The equation of motion for $g_{(1)ij}$ is 
\begin{eqnarray}
\label{oii}
\lefteqn{0 = \frac{1}{\ell} (g^{ij} g^{kl} g_{(1)kl}
-g^{ik} g^{jl} g_{(1)kl}) +\frac{\ell}{2}(R ^{ij}
-\frac{1}{2}g^{ij} R )} \nn
& & -\frac{\ell}{2}V (\frac{1}{2}g^{kl} g^{ij} 
-g^{ik} g^{jl} ) \partial_{k}\phi \partial_{l}\phi  
-\frac{\ell}{2}K (\frac{1}{2}g^{kl} g^{ij} 
-g^{ik} g^{jl} )
\partial_{k}\chi \partial_{l}\chi \nonumber \\
& & +\frac{\ell}{2}Q (\frac{1}{2}g^{kl} g^{ij} 
-g^{ik} g^{jl} )
\partial_{k}\chi \partial_{l}\phi  .
\end{eqnarray}
Here $V $, $K $, $Q $ are defined as follows:
\be
\label{oiii}
V  \equiv  X -Y ' \ ,\quad 
K  \equiv  Z -W ^{*} \ ,\quad 
Q  \equiv  W '+Y ^{*} \ .
\ee
The equations of motion for $\phi_{(1)}$ and $\chi_{(1)}$ are
\begin{eqnarray}
\label{oiv}
\lefteqn{0 = -\frac{4}{\ell}V \phi_{(1)}
+ \frac{2}{\ell}Q \chi_{(1)} 
-\frac{\ell}{2}V' g^{ij} \partial_{i}\phi 
\partial_{j}\phi } \nn
&& -\frac{\ell}{2}K' g^{ij} \partial_{i}\chi 
\partial_{j}\chi  
+\frac{\ell}{2}Q' g^{ij} \partial_{i}\phi 
\partial_{j}\chi \nonumber \\
& & +\frac{\ell}{\sqrt{-g }}\partial_{j}\{ \sqrt{-g }
V g^{ij}  \partial_{i}\phi \} 
-\frac{\ell}{2\sqrt{-g }}\partial_{j}\{ 
\sqrt{-g }Q g^{ij} \partial_{i}\chi \} \\
\lefteqn{0 = -\frac{4}{\ell}K \chi_{(1)}+ \frac{2}{\ell}Q 
\phi_{(1)} 
-\frac{\ell}{2}K^{*} g^{ij} \partial_{i}\chi 
\partial_{j}\chi }\nn 
&& -\frac{\ell}{2}V^{*} g^{ij} \partial_{i}\phi \partial_{j}\phi  
+\frac{\ell}{2}Q^{*} g^{ij} \partial_{i}\phi 
\partial_{j}\chi \nonumber \\
& & +\frac{\ell}{\sqrt{-g }}\partial_{j}\{ \sqrt{-g }
K g^{ij}  \partial_{i}\chi \} 
-\frac{\ell}{2\sqrt{-g }}\partial_{j}\{ 
\sqrt{-g }Q g^{ij} \partial_{i}\phi \}\ .
\end{eqnarray}
Then $g_{(1)ij}$, $\phi_{(1)}$, $\chi_{(1)}$ can be given 
in terms of $g_{ ij}$, $\phi $, $\chi $.
\begin{eqnarray}
\lefteqn{g_{(1)ij} = \frac{\ell^2}{2}\left(R_{ ij}-\frac{1}{6}
g_{ ij}R \right) + \frac{\ell ^2}{2}V \left\{\partial_{i}\phi 
\partial_{j}\phi -\frac{1}{6}g_{ ij}g^{kl} \partial_{k}\phi 
\partial_{l}\phi \right\} }\nonumber \\ 
& &+ \frac{\ell ^2}{2}K \left\{\partial_{i}\chi \partial_{j}
\chi  -\frac{1}{6}g_{ ij}g^{kl} \partial_{k}\chi 
\partial_{l}\chi \right\} 
- \frac{\ell ^2}{2}Q \left\{\partial_{i}\phi 
\partial_{j}\chi  -\frac{1}{6}g_{ ij}g^{kl} 
\partial_{k}\phi \partial_{l}\chi \right\} \label{gij1}\\ 
\lefteqn{\phi_{(1)} = \left\{ -\frac{\ell ^2}{16}
\left(\frac{Q \xi }{K V } -2\frac{V' }{V }\right)
g^{ij} \partial_{i}\phi \partial_{j}\phi 
-\frac{\ell^2}{16} \left( 2\frac{\eta }{V }
-\frac{Q K^{*} }{K V }\right)
g^{ij} \partial_{i}\chi \partial_{j}\chi \right.} \nn
&& \left. +\frac{\ell ^2}{8}\left(
\frac{K' Q }{K V } +2\frac{V^{*} }{V }\right)
g^{ij} \partial_{i}\phi \partial_{j}\chi  \right\}
\left( 1-\frac{1}{4}\frac{Q^{2} }{K V } 
\right)^{-1} +\frac{\ell ^2}{4}\triangle\phi \label{phi1} \\
\lefteqn{\chi_{(1)} = \left\{ -\frac{\ell ^2}{16}
\left(2\frac{\xi }{K }-\frac{Q V' }
{K V }\right)g^{ij} \partial_{i}\phi 
\partial_{j}\phi  \right. 
-\frac{\ell ^2}{16}\left( \frac{Q \eta }{K V }
-2\frac{K^{*} }{K }\right)
g^{ij} \partial_{i}\chi \partial_{j}\chi} \nn
& & \left. +\frac{\ell ^2}{8}\left(
\frac{V^{*} Q }{K V }
+2\frac{K' }{K }\right)
g^{ij} \partial_{i}\phi \partial_{j}\chi  \right\}
\left( 1-\frac{1}{4}\frac{Q^{2} }{K V } 
\right)^{-1} +\frac{\ell ^2}{4}\triangle\chi \label{chi1}  
\end{eqnarray}
Here $\eta$, $\xi$ are defined as follows:
\be
\eta  \equiv  K'  + Q^{*} =Z' +Y^{**}  \ ,
\quad 
\xi  \equiv V^{*} +Q' =X^{*} +W''  \ .
\ee 
For substituing (\ref{gij1}), (\ref{phi1}), (\ref{chi1}) into 
the action (\ref{Sln}), we rewrite (\ref{Sln}) as follows
\begin{eqnarray}
\lefteqn{S_{\ln } = \frac{1}{16\pi G} \int d^{4}x\sqrt{-g }\left\{
\frac{1}{2 \ell}g^{ij} g^{kl} 
\left(g_{(1)ij}g_{(1)kl}-g_{(1)ik}g_{(1)jl}\right)
\right.}\nonumber \\
& & +\frac{\ell}{2} \left(R ^{ij}-\frac{1}{2}g^{ij} R 
\right)g_{(1)ij} 
-\frac{2}{\ell}\{V \phi_{(1)}\phi_{(1)}+K \chi _{(1)}
\chi_{(1)} -Q \phi_{(1)} \chi_{(1)} \}\nonumber \\
& & -\frac{\ell}{2}V  \left(\frac{1}{2}g^{kl} g^{ij} 
-g^{ik} g^{jl} \right)
g_{(1)ij}\partial_{k}\phi \partial_{l}\phi  
-\frac{\ell}{2}K  \left(\frac{1}{2}g^{kl} 
g^{ij} -g^{ik} g^{jl} \right)
g_{(1)ij}\partial_{k}\chi \partial_{l}\chi  \nonumber \\ 
& & +\frac{\ell}{2}Q  \left(\frac{1}{2}g^{kl} 
g^{ij} -g^{ik} g^{jl} \right)
g_{(1)ij}\partial_{k}\phi \partial_{l}\chi 
-\frac{\ell}{2}\left( V' \phi_{(1)}+V^{*} \chi_{(1)} 
\right) g^{ij} \partial_{i}\phi \partial_{j}\phi  
\nonumber \\ 
& & -\frac{\ell}{2}\left( K^{*} \chi_{(1)}+K' \phi_{(1)} \right)
g^{ij} \partial_{i}\chi \partial_{j}\chi 
+\frac{\ell}{2}\left( Q' \phi_{(1)}+Q^{*} \chi_{(1)} \right)
 g^{ij} \partial_{i}\phi \partial_{j}\chi  \nonumber \\ 
& & -\ell V g^{ij} \partial_{i}\phi 
\partial_{j}\phi_{(1)} -\ell K g^{ij} \partial_{i}\chi 
\partial_{j}\chi_{(1)} \left.-\frac{\ell}{2}Q 
\left(g^{ij} \partial_{i}\phi_{(1)}\partial_{j}\chi  
+g^{ij} \partial_{i}\phi \partial_{j}\chi  
\right)\right\} \nonumber 
\end{eqnarray}
Substituing (\ref{gij1}), (\ref{phi1}), (\ref{chi1}) into this 
action, we find 
\begin{eqnarray}
\lefteqn{S_{\ln } = \frac{\ell ^3}{16\pi G} \int d^{4}x
\sqrt{-g }\left\{ \frac{1}{8}R^{ij} R_{ ij}
-\frac{1}{24}R ^{2}\right.}\nonumber \\
& &+\frac{V }{4}\left( R^{ij} \partial_{i}\phi 
\partial_{j}\phi  -\frac{1}{3}R g^{ij} 
\partial_{i}\phi \partial_{j}\phi  \right) 
+\frac{K }{4}\left( R^{ij} \partial_{i}\chi 
\partial_{j}\chi  -\frac{1}{3}R g^{ij} 
\partial_{i}\chi \partial_{j}\chi  \right) \nonumber \\
& &-\frac{Q }{4}\left( R^{ij} \partial_{i}\phi 
\partial_{j}\chi  -\frac{1}{3}R g^{ij} 
\partial_{i}\phi \partial_{j}\chi  \right) 
-\frac{1}{12} \left\{ (V g^{ij} 
\partial_{i}\phi \partial_{j}\phi )^2
+(K g^{ij} \partial_{i}\chi \partial_{j}\chi )^2 \right. \nn
&& \left. + (Q g^{ij} \partial_{i}\phi 
\partial_{j}\chi )^2 \right\} 
-\frac{K V }{4}\left\{ 
\frac{1}{3}(g^{ij} \partial_{i}\phi \partial_{j}\phi )
(g^{kl} \partial_{k}\chi \partial_{l}\chi )
-(g^{ij} \partial_{i}\phi \partial_{j}\chi )^2 \right\} \nn
&& -\frac{V Q }{6}(g^{ij} \partial_{i}\phi \partial_{j}\phi )
(g^{kl} \partial_{k}\phi \partial_{l}\chi ) 
-\frac{K Q }{6}(g^{ij} \partial_{i}
\chi \partial_{j}\chi )
(g^{kl} \partial_{k}\chi \partial_{l}\phi ) \nn
&& + \left[ \frac{1}{32}
\left( \frac{{V' }^2 }{V }+\frac{\xi ^2 }{K }- 
\frac{V' Q \xi }{V K }\right)
(g^{ij} \partial_{i}\phi \partial_{j}\phi )^2 \right. \nn
&& +\frac{1}{32}\left( \frac{{K^{*} }^2 }{K }
+\frac{\eta ^2 }{V }- 
\frac{K^{*} Q \eta }{V K }\right)
(g^{ij} \partial_{i}\chi \partial_{j}\chi )^2 \nn
&& +\frac{1}{8}\left( \frac{{K' }^2 }{K } +\frac{{V^{*} }^2 }{V }
-\frac{V^{*} Q K' }{V K }\right)
(g^{ij} \partial_{i}\phi \partial_{j}\chi )^2 
\nonumber \\
& &+\frac{1}{16} \left( -\frac{K^{*} }{K }\xi 
-\frac{V' }{V }\eta 
+\frac{Q }{2V K }(K^{*} V' 
+\xi \eta ) \right) (g^{ij} \partial_{i}\phi \partial_{j}\phi )
(g^{kl} \partial_{k}\chi \partial_{l}\chi ) 
\nonumber \\
& &+\frac{1}{16}\left( -2\frac{K' }{K }\xi 
+2\frac{V' V^{*} }{V }\eta 
+\frac{Q }{V K }
(K' V' -V^{*} \xi ) \right) 
(g^{ij} \partial_{i}\phi \partial_{j}\phi )
(g^{kl} \partial_{k}\phi \partial_{l}\chi ) 
\nonumber \\
& & \left. +\frac{1}{16}\left( -2\frac{V^{*} }{V }\eta 
+2\frac{K' K^{*} }{K }\eta 
+\frac{Q }{V K }
(K^{*} V^{*} -K' \eta ) \right)
(g^{ij} \partial_{i}\phi \partial_{j}\chi )
(g^{kl} \partial_{k}\chi \partial_{l}\chi )\right] \nn
&& \times \left(1-\frac{Q ^2 }{4K V } 
\right)^{-1} 
+\frac{V }{8}\left( \triangle\phi \right)^{2} 
+\frac{K }{8}\left( \triangle\chi \right)^{2} 
-\frac{Q }{4}\triangle\phi \triangle \chi \nn
&& -\frac{1}{16}\left[ \left(\frac{Q }{K }\xi 
-2V' \right)g^{kl}\partial_{k}\phi \partial_{l}\phi 
\triangle\phi \right. 
+\left(\frac{Q }{V }\eta 
-2K^{*}  \right)g^{kl}\partial_{k}\chi 
\partial_{l}\chi \triangle\chi \nn
&& -\frac{Q }{2V }\left(\frac{Q }{K }\xi 
-2V' \right) g^{kl}\partial_{k}\phi \partial_{l}\phi 
\triangle\chi \nonumber \\
& & \left. -\frac{Q }{2K }\left(
\frac{Q }{V }\eta  -2K^{*}  \right)
g^{kl}\partial_{k}\chi \partial_{l}\chi \triangle\phi \right]  
\times \left(1-\frac{Q ^2 }{4K V } \right)^{-1} 
\nonumber \\
& & +\frac{V^{*} }{4}g^{kl}\partial_{k}\phi 
\partial_{l}\chi \frac{1}{\sqrt{-g }}
\partial_{j}\left(\sqrt{-g }g^{ij} 
\partial_{i}\phi  \right) \left.+\frac{K' }{4}g^{kl}\partial_{k}
\phi \partial_{l}\chi \triangle\chi \right\}
\end{eqnarray}
If we define 
\be
\label{Ni}
\left(\begin{array}{c}X^1 \\ X^2 \\ \end{array}\right)
\equiv \left(\begin{array}{c}\phi \\ \chi \\ \end{array}\right)
\ ,\quad h_{\mu\nu}\equiv \left(\begin{array}{cc}V &-{Q \over 2} \\ 
-{Q \over 2} &K \\ \end{array}\right)\ ,
\ee
we obtain
\be
\label{Nii}
h\equiv \det h_{\mu\nu}=VK-{Q^2 \over 4}\ ,\quad 
h^{\mu\nu}={1 \over h}
\left(\begin{array}{cc} K & {Q \over 2} \\ 
{Q \over 2} & V \\ \end{array}\right)\ .
\ee
Then Eqs.(\ref{gij1}), (\ref{phi1}) and (\ref{chi1}) 
can be rewritten as follows:
\bea
\label{Niii}
g_{(1)ij}&=&{l^2 \over 2}\left\{\left(R_{ij} 
- {1 \over 6}g_{ij}R\right) 
+ h_{\mu\nu}\left(\partial_i X^\mu \partial_j X^\nu
-{1 \over 6}g_{ij}\left(\partial X^\mu\cdot 
\partial X^\nu\right)\right) \right\} \nn
\label{Niv}
X_{(1)}^\mu&=&{l^2 \over 4}\Gamma^{\mu}_{\rho\sigma}
\left(\partial X^\rho\cdot 
\partial X^\sigma\right) + {l^2 \over 4}\triangle X^\mu \ .
\eea
Here 
\be
\label{Nv}
\left(\partial X^\mu\cdot \partial X^\nu\right)
\equiv g^{ij}\partial_i X^\mu \partial_j X^\nu 
\ee
and $\Gamma^\mu_{\nu\rho}$ is a connection on the 
target manifold:
\be
\label{Nvi}
\Gamma^\mu_{\nu\rho}={1 \over 2} h^{\mu\tau}\left(h_{\nu\tau,\rho}
+ h_{\rho\tau,\nu} - h_{\nu\rho,\tau}\right)\ .
\ee
On the other hand, $S_{\ln}$ can be rewritten as follows
\bea
\label{Nvii}
\lefteqn{S_{\ln}={1 \over 16\pi G}\int d^4 x \sqrt{-g}
\left\{ {1 \over 2l}g^{ij}g^{kl}
\left(g_{(1)ij}g_{(1)kl} - g_{(1)ik}g_{(1)jl}\right) 
\right. } \nn
&& + {l \over 2}\left(R^{ij}-{1 \over 2}R g^{ij}\right)
g_{(1)ij} - {2 \over l}h_{\mu\nu}X_{(1)}^\mu X_{(1)}^\nu
\nn
&& - {l \over 2}h_{\mu\nu}
\left({1 \over 2}g^{kl}g^{ij} - g^{ik}g^{jl}\right)
g_{(1)ij}\partial_k X^\mu \partial_l X^\nu \nn
&& \left.-{l \over 2}h_{\mu\nu,\rho}X_{(1)}^\rho 
g^{ij}\partial_i X^\mu \partial_j X^\nu 
-l h_{\mu\nu}g^{ij}\partial_i X^\mu 
\partial_j X_{(1)}^\nu\right\}\ .
\eea
Substituting (\ref{Niii}) and (\ref{Niv}) into 
(\ref{Nvii}) and integrating by parts, we obtain
\bea
\label{Nviii}
\lefteqn{S_{\ln}={1 \over 16\pi G}\int d^4 x \sqrt{-g}
\left[l^3\left(-{1 \over 24}R^2 
+ {1 \over 8}R_{ij}R^{ij}\right) \right. } \nn
&& + {l^3 \over 4}R^{ij}h_{\mu\nu}
\partial_i X^\mu \partial_j X^\nu 
- {l^3 \over 12}R h_{\mu\nu} 
\left(\partial X^\mu\cdot \partial X^\nu\right) \\
&& -{l^3 \over 24}\left\{ h_{\mu\nu} \left(
\partial X^\mu\cdot \partial X^\nu\right)\right\}^2
+{l^3 \over 8}h_{\mu\nu}h_{\rho\sigma}
\left(\partial X^\mu\cdot \partial X^\rho\right)
\left(\partial X^\nu\cdot \partial X^\sigma\right) \nn
&& \left. + {l^8 \over 8}h_{\mu\nu}
\left\{\triangle X^\mu + \Gamma^\mu_{\rho\sigma}
\left(\partial X^\rho\cdot \partial X^\sigma\right)
\right\}
\left\{\triangle X^\nu + \Gamma^\nu_{\tau\eta}
\left(\partial X^\tau\cdot \partial X^\eta\right)
\right\}\right] \ .\nonumber
\eea
If we put 
\be
\label{Nix}
h_{\mu\nu}=\left(\begin{array}{cc}
2 & 0 \\ 0 & c \\ \end{array} \right)\ , \ 
(c\ \mbox{is an arbitrary constant}),\ 
\quad X^2=0
\ee
the previous result in \cite{NOa} for dilatonic gravity can be 
reproduced. 

If we choose the action as motivated by the bosonic sector of
type IIB supergravity with RR-scalar \cite{GGP}
\begin{eqnarray}
\label{ov}
S = \frac{1}{16\pi G} \int _{M_{d+1}} 
d^{d+1}x \sqrt{-\hat{G}}
\left( {\hat R} + \frac{1}{2} (\hat{\nabla } \phi)^{2} 
+ \frac{1}{2}\e^{2\phi}(\hat{\nabla } \chi)^{2} \right). 
\end{eqnarray}
the functions $X(\phi,\chi)$, $Y(\phi,\chi)$, $Z(\phi,\chi)$,
$W(\phi,\chi)$ become 
\be
\label{ovi}
X(\phi,\chi) = + \frac{1}{2} \ , \quad
Z(\phi,\chi) = \frac{1}{2}\e^{2\phi} \ , \quad
Y(\phi,\chi) = W(\phi,\chi)=0 \ .
\ee
Then the $\ln\epsilon$ terms of the action, corresponding to the 
anomaly, are written in following form.
\bea
\label{ovii}
\lefteqn{S_{\ln } = \frac{\ell^3}{16\pi G} \int d^{4}x
\sqrt{-g }\left\{
\frac{1}{8}R^{ij} R_{ij}-\frac{1}{24}R ^{2}
\right.}\nonumber \\
& & -\frac{1}{8}\left( R^{ij} \partial_{i}\phi 
\partial_{j}\phi  -\frac{1}{3}R g^{ij} 
\partial_{i}\phi \partial_{j}\phi  \right)
+\frac{1}{8}\e^{2\phi }
\left( R^{ij} \partial_{i}\chi \partial_{j}\chi 
-\frac{1}{3}R g^{ij} \partial_{i}\chi 
\partial_{j}\chi  \right) \nonumber \\
& & -\frac{1}{48}
\left\{ (g^{ij} \partial_{i}\phi \partial_{j}\phi )^2
+\e^{4\phi }(g^{ij} \partial_{i}\chi 
\partial_{j}\chi )^2 
\right\} \nonumber \\
& &+\frac{1}{16}\e^{2\phi }\left\{ 
\frac{1}{3}(g^{ij} \partial_{i}\phi \partial_{j}\phi )
(g^{kl} \partial_{k}\chi \partial_{l}\chi )
-(g^{ij} \partial_{i}\phi \partial_{j}\chi )^2 
\right\} \nonumber \\
& &-\frac{1}{16}\e^{4\phi }(g^{ij} 
\partial_{i}\chi \partial_{j}\chi )^2 
+\frac{1}{4}\e^{2\phi }
(g^{ij} \partial_{i}\phi \partial_{j}\chi )^2 
\nonumber \\
&& \left. -\frac{1}{16}\left(\triangle\phi\right)^{2} 
+\frac{1}{16}\e^{2\phi }\left(\triangle\chi \right)^{2} 
+\frac{1}{4}\e^{2\phi }g^{kl}\partial_{k}\phi 
\partial_{l}\chi\triangle \chi \right\}\ .
\eea
The Weyl anomaly coming from the multiplets of ${\cal N}=4$ 
supersymmetric $U(N)$ or $SU(N)$ Yang-Mills 
conformally coupled with ${\cal N}=4$ 
conformal supergravity was calculated in 
\cite{HT}:\footnote{See Eqs.(2.5) and (2.6) in \cite{HT}.}
\bea
\label{xxi}
T&=&-{N^2 \over 4(4\pi)^2}\left[2\left(R_{ij}R^{ij}
-{1 \over 3}R^2\right)+F^{ij}F_{ij} \right. \\
&& \left. + 4\left\{ 2\left( R^{ij} - {1 \over 3} Rg^{ij}\right)
\partial_i\varphi^*\partial_j\varphi  
+\left|\triangle\varphi \right|^2 \right\} + \cdots \right]\ .
\nonumber
\eea
Here $F_{ij}$ is the field strength of SU(4) gauge fields, $\varphi$ 
is a complex scalar field which is a combination of dilaton and 
RR scalar and ``$\cdots$'' expresses the terms containing other 
fields in ${\cal N}=4$ conformal supergravity multiplet and 
higher powers of the fields. It is also interesting that above conformal 
anomaly actually gives the (bosonic sector) of ${\cal N}=4$ conformal SG.

If we choose
\be
\label{xx}
{l^3 \over 16\pi G}={2N^2 \over (4\pi)^2}\ , \quad 
\varphi = \phi + \e^\phi \chi\ , \quad 
\varphi^* = -\phi + \e^\phi \chi
\ee
and consider the background where only gravity and the 
scalar field $\varphi$ in the ${\cal N}=4$ conformal supergravity 
multiplet are non-trivial and other fields vanish in (\ref{xxi}), 
Eq.(\ref{ovii}) exactly reproduces the result in (\ref{xxi}).

Thus, we got axion-dilatonic Weyl anomaly from SG side. It is remarkable 
that it coincides with perturbative result even in presence of axion, not
only 
in dilaton-gravitational background. 

It is not difficult to get the corresponding result in other dimensions (we 
discussed only d4 case). It would be of interest also to extend our results 
for gauged SGs. To do this one should add the axion-dilatonic 
(or in minimal case, only dilatonic) potential to initial action. 
This may help in construction of new versions of multidimensional 
gauged SGs.
In its own turn, the analysis like above may also be useful in the 
extending of AdS/CFT correspondence to gauged SGs.

\end{document}